# Highly efficient exciton-exciton annihilation in single conjugated polymer chains

Nicola J. Fairbairn, Olga Vodianova and Gordon J. Hedley*

School of Chemistry, University of Glasgow, Glasgow, G12 8QQ, United Kingdom

*Email: Gordon.Hedley@glasgow.ac.uk Telephone: +44 (0)141 330 1763

**Abstract**

The number of excitons that conjugated polymers can support at any one time underpins their optoelectronic performance in light emitting diodes and as laser gain media, as it sets a natural limit on exciton density. Here we have measured the time-resolved photon statistics of single chains of polyfluorene to extract the absolute number of independent emitting sites present and its time dependence. We find that after 100 ps each chain can only support 1 or 2 independent excitons, and that even at the earliest times this number rises only to 4, suggesting a high degree of electronic coupling between chromophores that facilitates efficient exciton-exciton annihilation. In circumstances where a low density of low-energy sites is present, annihilation between them still dominates. The results indicate that achieving high exciton densities in conjugated polymers is difficult, and in applications where it is desirable new strategies should be devised to control exciton-exciton annihilation.

Conjugated polymers have attractive optoelectronic properties for use in organic light emitting diodes (OLEDs), photovoltaic cells and lasers.[1–10] Excited states are typically Frenkel excitons, delocalised coulombically-bound electron-hole pairs. These excitons hop via Förster resonance energy transfer (FRET) initially downhill, then within the available thermal energy via diffusion.[10,11] How many excitons a polymer chain can support simultaneously, its exciton density, is important in defining its overall optoelectronic performance as it sets a limit on relevant device parameters, e.g. OLED brightness. Exciton motion via FRET enables two of them to come in close spatial proximity to each other, where they can undergo exciton-exciton annihilation (EEA),[12–16] a form of self-FRET that leads to the loss of one exciton per encounter event. Consequently, for mobile excitons, annihilation determines the upper limit of exciton density, and thus is important to understand and control in organic semiconducting devices, such as in an OLED operating under high current injection for high brightness, where the overall efficiency of the device decreases,[17] or in an organic laser, where annihilation is a loss channel that increases thresholds and works against any attempt at electrical injection.[18]

Measuring the absolute upper limit of allowable exciton densities in conjugated polymers and how this evolves during the excited state lifetime due to annihilation is inherently difficult. An indication that the density limit has locally been reached in some places is when annihilation is observed. This requires high photon fluences when under optical excitation i.e. a significant fraction of the excitons need to be annihilating for this to be measurable in photoluminescence or transient absorption.[12–16] However, an underlying non-annihilating population typically remains that needs to be divided out, complicating analysis. This is exacerbated when considering that non-uniform laser excitation profiles (normally Gaussian) are used. Secondly, any attempt to determine the absolute exciton density requires assumptions on the mass density and number of optically active chromophores in the sample volume that is being measured. The former is non-trivial to measure accurately, while the latter is subject to significant error.

Here we have used single molecule spectroscopy[19–21] to study individual chains of polyfluorene. By measuring the time-resolved photon statistics of fluorescence from single polymer chains for the first time we can automatically monitor exciton-exciton annihilation in the very rare cases where it happens, free from the much larger background of circumstances where it does not – in our work this is at a rate of 35 per million emission events. We find that annihilation is always present as a loss channel. We observe that each individual polyfluorene chain is comprised of only ~4 independent emitting sites at the earliest times we can measure (~50 ps), suggesting that fast sub-picosecond evolution of the excited state significantly limits how many excitons a single chain can support. This number drops to ~2 within 100 ps, i.e. a single polymer chain can only support two excitons or less after this time. Exploiting keto-defects in polyfluorene as an analogy for exciton protection strategies (low energy sites surrounded by protective higher energy ones) we find that these low energy sites can still annihilate with each other, with 5 independent sites becoming 2 after 1 ns. Our findings suggest enabling high exciton densities in conjugated polymers will require careful control of energetic and conformational landscapes to enable applications such as high brightness OLEDs or electrical injection organic lasers.

**Results**

Poly(9,9'-dioctylfluorene) (PFO), chemical structure inset Figure 1a, is a prototypical conjugated polymer which has been widely studied owing to its desirable deep blue emission and high photoluminescence (PL) quantum yield.[22–25] Optical or electrical excitation of PFO leads to the formation of excitons delocalised across 4-5 monomeric units.[26] Single chains of PFO embedded in a PMMA matrix were measured in a nitrogen environment on a homebuilt confocal fluorescence microscope with single photon counting, Figure 1b (see Supplementary Information for a full description of experimental and sample preparation methods). Fluorescence scan images over a region were recorded to identify single chains of PFO, before each chain was then measured for longer (up to 30 second) durations. Expected single-chain PFO behaviour is observed,[27–33] including fast mono-exponential lifetimes, Figure 1c, and intensity blinking (see Supplementary Information). When emitted photons are measured on two detectors with a 50:50 beamsplitter in a Hanbury Brown and Twiss arrangement[34] photon antibunching[35] is observed, Figure 1d. Here the correlation events, N, between the two detector channels are recorded for different time lags of laser pulse periods (Δt). The correlation between photons detected at their "true" time (Δt = 0) is calculated ($N_C$) along with at artificial time lags (Δt ≠ 0), ($N_L$). The ratio between these is used to determine the number of independent emitting sites, $n$, on the chain using equation 1.[36]

$$n = \frac{1}{\left(1 - \frac{N_C}{N_L}\right)} \qquad \text{(equation 1)}$$

For 2215 measured chains of PFO, $n$ is found to be 2.07 ± 0.09. Crucially this is a time-averaged result, i.e. over the excited state lifetime there are 2.07 active independent emitting sites present in each PFO chain, weighted to early time where more photons are emitted. It is then pertinent to ask how this will vary with time.

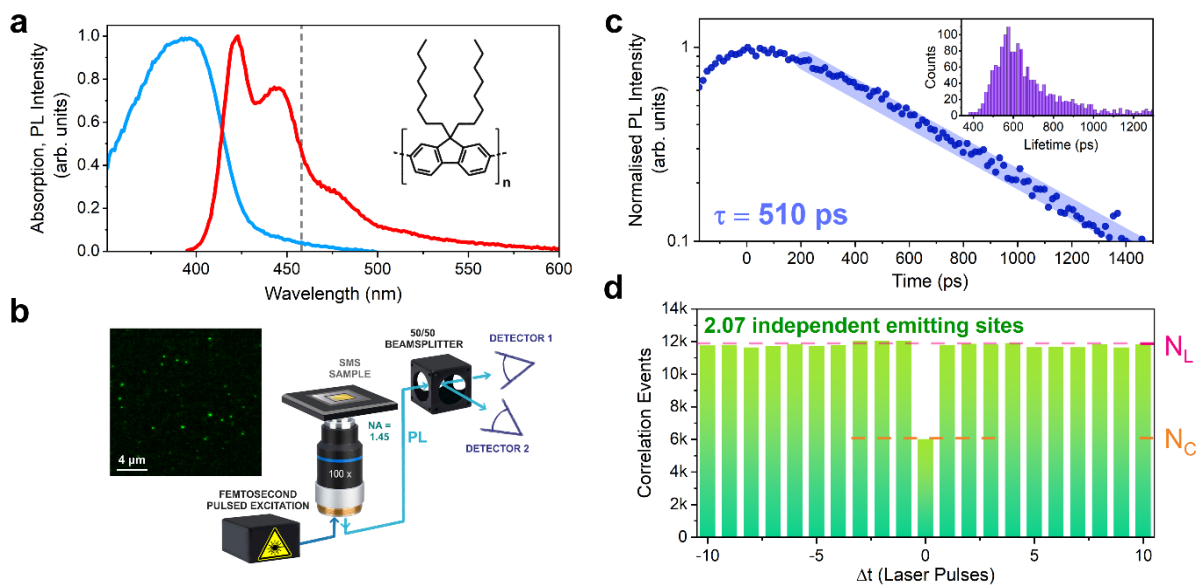

**Figure 1: (a)** Absorption and photoluminescence spectra of poly(9,9'-dioctylfluorene) (PFO), the inset chemical structure shows a monomeric unit of PFO. **(b)** Schematic of experimental fluorescence confocal microscopy setup used to measure single polymer chains, with a Hanbury Brown and Twiss geometry for photon correlation measurements. A typical fluorescence scan image is shown with emission from individual spatially isolated chains. **(c)** Photoluminescence decay of a single PFO chain, with a fitted lifetime of 510 ps, inset is a histogram of lifetimes for 1821 individual PFO chains. **(d)** Photon antibunching histogram obtained from 2215 individual PFO chains. The number of correlation events in the central bin at Δτ = 0 is defined as $N_C$, and the mean number of correlation events in the lateral time-lagged bins as $N_L$. The ratio $N_C/N_L$ allows determination of the number of independent emitting sites, *n*, here calculated to be 2.07 ± 0.09.

To explore any time dependence on the number of emitting sites we then apply our recently developed time-resolved antibunching (TRAB) technique[37] to single polymer chains for the first time. To test that there is a response we first applied three large filter windows, Figure 2a, to the recorded photons, yielding separate antibunching results for each respective filter window, Figure 2b. Evident immediately is that there is a change in the number of independent emitting sites on single chains of PFO between these time windows. At the earliest times (0-150 ps), ~2.7 are present. In the next (160-360 ps) this drops to ~2.1, while in the final window (370-870 ps) this drops further to ~1.7. We have adjusted the lengths of the windows to ensure there are similar numbers of correlation events (~4000 on lateral peaks) in each, and thus similar signal-to-noise ratios.

Next, rather than looking at defined large windows we can instead calculate a continuous measure of the ratio between the central and lateral values ($N_C/N_L$), with a step size of 20 ps, as shown in Figure 2c, for 2215 individually measured chains. This can then be converted into the more meaningful number of independent emitting sites via equation 1, which is plotted in Figure 2d. Now, a direct readout of how many sites are active in the emission process in single isolated chains of PFO can be made across its excited state lifetime. What is found is that there is a reduction in this number, from ~4 to 1.6, over the first 600 ps, but then little further change out to at least 1 ns. This reduction directly tells us that excitons are interacting with each other on single chains. The most probable explanation for this is exciton-exciton annihilation. The timescale over which annihilation proceeds (0-600 ps) is suggestive that it is facilitated by exciton motion along the chain, and either two singlet excitons meet and annihilate, or one moves and finds a relatively static second (low energy trap site). Consequently, the results presented here offer an ability to observe on-chain exciton annihilation free from any background population of non-annihilating excitons. To quantify this, we can compare the number of occurrences where two photons are detected after excitation ($N_C$) with the events where only one photon is only detected. We find that this is 35 per million detected photons, i.e. the measured on-chain annihilation events in PFO are very rare and do not lead to any deviation in the conventionally observed photophysical properties (PL decay etc.), yet they do exist, and we are uniquely sensitive to them with this measurement.

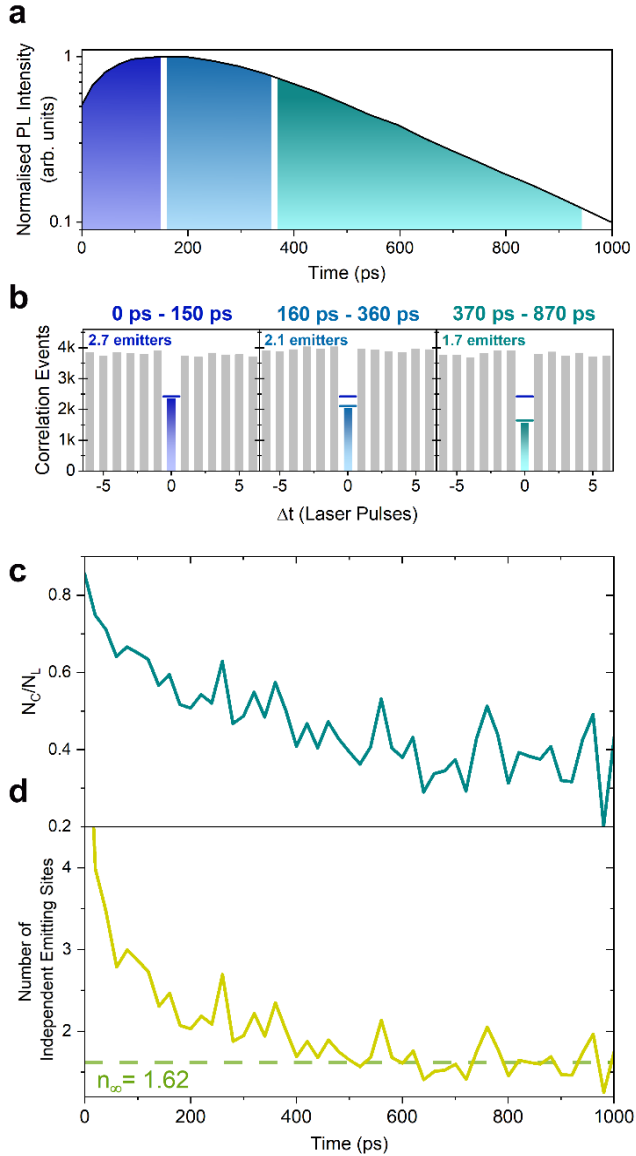

**Figure 2: (a)** The measured photoluminescence decay of PFO with the shaded regions indicating the photon arrival time windows used to construct corresponding antibunching histograms shown in **(b)**. In the early window (0 to 150 ps), the dip in the central bin indicates 2.7 emitting sites, dropping to 2.1 in the 160 to 360 ps window, before dropping further to 1.7 emitters in the final window of 370 to 870 ps. The $N_C$ level for the first window is shown on all three plots to evidence the drop. **(c)** Plot of the continuously calculated $N_C/N_L$ ratio over the first 1 ns after the laser pulse with a resolution of 20 ps, showing the drop in the ratio. This is more comprehensible if the values are converted using equation 1 into the number of independent emitting sites, as shown in **(d)**, with a drop from ~4 to ~1.6 over the first 600 ps.

Measuring single PFO chains provides a rare opportunity to extract valuable information on excited state process not readily accessible with ensemble techniques. In results presented so far, we have neglected any consideration of heterogeneity in the chains, treating them as all behaving the same way. Many single molecule spectroscopy studies indicate this is seldom the case for conjugated polymers.[38–44] To investigate this we have combined a conventional Hanbury Brown and Twiss setup with a 3rd detector that splits emission with a dichroic mirror centred at 458 nm, Figure 3a. Ultimately, this introduces a means of measuring spectral fluctuations in a single PFO chain whilst simultaneously allowing recording spectrally unbiased photon correlations. Now we can also monitor the spectral position of emission from PFO chains and selectively filter the TRAB correlation results for different circumstances. To allow assessment of how the spectral position of emission changes we define a value of merit, the colour ratio:

$$Colour\ Ratio = \frac{I_{\lambda>458}-I_{\lambda<458}}{I_{\lambda>458}+I_{\lambda<458}} = \frac{C_1-C_2}{C_1+C_2} \qquad \text{(equation 2)}$$

where $C_1$ and $C_2$ are the recorded counts on the two detectors split by the dichroic mirror. Negative colour ratio values indicate emission dominated by λ < 458 nm, associated with ensemble steady state PFO emission, while positive values indicate redder emission, potentially associated with the fluorenone keto-defect in PFO.[45–48] Plotting the colour ratio over 30 seconds for a typical single chain, Figure 3b, we observe substantial variation. The most common value is ~-0.2, consistent with -0.24 obtained by simulating transmission of a measured ensemble steady state PFO emission spectrum through the setup (see Supplementary Information for a full description of the simulation). However, occasional and temporary jumps in the colour ratio are also observed, up to values close to 1, consistent with emission by fluorenone, with simulated expected colour ratio values for it emitting alone being 0.945 (spectrum from [48], see Supplementary Information for details). These jumps are, however, fully reversible, indicating that conventional PFO emission is recoverable, even after fluorenone has been detected on the chain. This can be explained when one considers that even one or two keto-defects on a chain can give rise to fluorenone emission,[49] but whether the exciton reaches and emits from them depends on where the exciton forms and whether it can hop to that site. The constantly changing colour ratio we observe suggests the formation of long-lived species can shut down sections of the polymer chain from absorbing/emitting, thus enabling (or forcing) excitons to emit from other sites for some period of time. Typically, these dynamics are challenging to disentangle as the blocking species are non-emissive, so information about which excitonic sites are active is hard to obtain. Here the keto-defects act as low-energy trap sites on the chain, and emission from them is spectrally separable from PFO,[50] thus detection of their presence and effects can be used to determine exciton behaviour on single chains.

Sampling all measured values of colour ratio for 5012 chains with 20 ms binning with at least 200 counts present (see Supplementary Information for a full description of this conditional method) we can construct a histogram of the population, as shown in Figure 3c. The expected values for emission solely from PFO or fluorenone, as noted above, are shown as dashed lines, and two peaks in the histogram are observed, consistent with these two species. In general, energetic heterogeneity is expected when measuring single chains, so the broad distribution centred on -0.2 likely represents a variety of conformations of polyfluorene. The fluorenone peak is narrow, however this is most likely caused by the limited spectral discrimination due to the dichroic mirror's central wavelength being far away from this emission. Existence of a population between the two peaks (shaded yellow) is not entirely unexpected and is because when we sample over 20 ms in this histogram region we may have a mixture of PFO and fluorenone emission present, giving an average value between the two. This explanation is favoured over the presence of any red-shifted PFO such as beta phase segments,[27] as they give a simulated colour ratio of 0.008 from their steady state PL spectrum, a value significantly lower than the range shaded yellow here.

As noted above, these spectrally-resolved single chain measurements have been made with three detectors. Merging the recorded photons on detectors 1 and 2 in software post-measurement, we create a new virtual detector, that can then be correlated with detector 3 for exploration of time-resolved photon antibunching with spectral sensitivity. Three highlighted regions from the colour ratio histogram have been chosen for this, blue for PFO dominant (-0.6 to 0.1), pink for fluorenone (0.8 to 1.0) and yellow for mixed (0.2 to 0.7). In software analysis routines (see Supplementary Information for a fuller description), we filter for periods of time when these colour ratios are present and then perform time-resolved photon antibunching, as previously described. The calculated number of independent emitting sites present versus time for the three highlighted regions are shown in Figure 3d – we note that if no spectral filtering is applied, TRAB results almost identical to those presented in Figure 2 can be extracted (see Supplementary Information). In considering these spectrally filtered time-resolved antibunching results it is important to establish what we are measuring. When a specific spectral region is selected then we are sensitive to excited states in that energy window interacting with, and only with, each other, i.e. any interactions/losses with excitons of energies outside the filter window will not be registered. Immediately clear is that when chains have PFO or mixed PFO-fluorenone emission they behave similarly, with a fast drop in the number of emitting sites, while chains that solely have fluorenone emission show a slower decay. Describing each area in turn: when emission is from PFO alone (Figure 3d, blue data, solid line) they show fast reduction in the number of available emitting sites, in contrast to the spectrally unresolved results in Figure 2. We are left with only 2 independent emitting sites present on large (average $M_w \geq 200$ kDa, $M_n = 54$ kDa) polymer chains after only 100 ps. The time-resolution here restricts full exploration of the early time behaviour, but at the earliest we can resolve there are only 3 or 4 emitting sites. A small slower decay in the number of emitters (inset Figure 3d) is observable and is consistent with on-chain exciton-exciton annihilation, but in general very little annihilation takes place after 100 ps when polyfluorene chromophores alone are emitting. In contrast, for the periods of time when the chain is emitting from a fluorenone site (Figure 3d, red data, dashed line), the number of independent emitting sites begins high (~5) and decays slowly, reaching only 2 after 1 ns. As noted above, in these spectral measurements we are sensitive only to the interaction we filter for, i.e. here we are uniquely only going to observe how excited states situated on fluorenone interact with each other. In a similar way, the region intermediate between PFO and fluorenone, (Figure 3d, yellow data, dotted line), primarily represents encounter events between PFO and fluorenone excited states. This decay in the number of independent emitting sites follows the PFO only one closely, consistent with mobile PFO excitons defining the overall interaction.

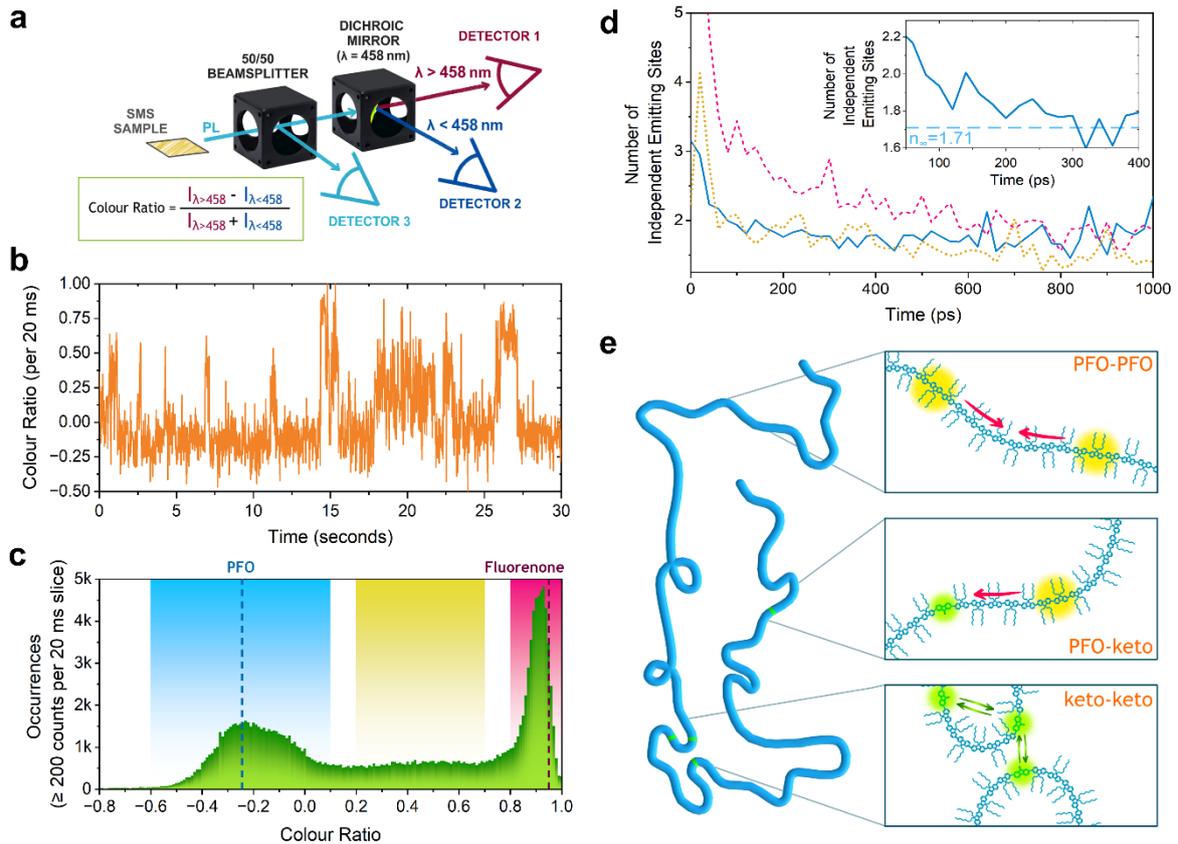

**Figure 3: (a)** Optical detection scheme enabling spectrally filtered Hanbury Brown and Twiss measurements. Detectors 1 and 2 can be used to define regions of time with a specific colour ratio (equation inset), derived from photons split on the dichroic mirror between them. Counts on these two detectors can then be summed and correlated with those on detector 3 for this region of time. **(b)** A single PFO chain time trace measured over 30 s, exhibiting significant fluctuations in colour ratio ranging from -0.25 to 1. **(c)** Histogram of colour ratio values per 20 ms bin with at least 200 counts present, for 5012 chains. Simulated colour ratios from steady state PFO (-0.24, blue vertical dashed line) and fluorenone (0.945, pink vertical line) emission spectra are also shown along with larger shaded regions. **(d)** Time-resolved photon antibunching derived plot of the number of independent emitting sites as a function of time for the three shaded spectral regions of colour ratio in panel c, with a zoomed-in view inset to aid clarity. **(e)** Schematic of a single polymer chain, with the three observed scenarios: blue excitons annihilating with each other (top), blue annihilating with keto defects (middle) and keto defects annihilating with each other (bottom).

We now turn our attention to the interpretation of these results, Figure 3e. For PFO excitons interacting with each other, the loss of independent emitting sites is remarkably fast, close to complete in 100 ps along an entire conjugated polymer chain. The losses we see are real number losses, i.e. represent exciton-exciton annihilation and a reduction in the ability for the single chain to host excitons and emit from multiple sites simultaneously. It is likely that the single chains we are measuring are not straight, so some pseudo-interchain coupling across different segments of the same chain will be present, enhancing exciton motion and annihilation. Our results have significant implications in defining limitations on light emission in PFO. Coupling between chromophoric sites is very strong at times < 100 ps, leading to fast loss of population and ensuring a single chain is only able to support one or sometimes two excitons after this time. This puts an upper limit on how many photons can be emitted, whether through optical excitation or charge injection. In thinking about how to support higher exciton densities, structures that provide an energetic barrier surrounding an emitting centre to prevent motion/interaction between them is attractive. This can conveniently be tested with the fluorenone keto-defects present in PFO, where we can exclusively look at how they interact with each other. Surprisingly, despite very likely only being a few[49] defect sites per chain spaced apart from each other, these sites interact, with a loss from 5 down to ~2 occurring, but over 1 nanosecond. As these are low-energy sites, it is anticipated that the barrier of surrounding chromophores with higher energy will ensure excitons are relatively trapped (from the PL peaks for both, this should be a barrier of ~0.5 eV). Consequently, we suggest that a combination of longer-range slower FRET, and circumstances where two fluorenone sites are close (either because they happen to be next to each other, or more likely across chain segments) explains the observed behaviour. This implies that attempts to create polymer chains with "protected" emitting sites using energy barriers will not easily work, as longer-range transfer and interchain pathways will generally ensure annihilation still occurs.

**Discussion**

In this work we have observed very rare circumstances where exciton-exciton annihilation on a single conjugated polymer chain occurs for the first time. We find that these events are ~35 per million, i.e. we are measuring in a regime far away from high exciton densities which are normally required to see annihilation. We find that excitons on chromophores consistent with conventional PFO emission interact with each other very readily. Three or four independent sites are present on the chain at the earliest times we can detect (~50 ps), and this reduces to ~2 in just 100 ps. When observing fluorenone sites annihilating with each other there is still a reduction down to ~2, but this takes ~1 ns. Consequently, these results suggest a fundamental limit in materials with a high degree of exciton coupling to support large exciton densities, particularly when our observations are the best-case scenario, free from the very significant interchain interactions that would be present in mesoscopic and bulk regimes found in devices.

**Author Contributions**

N.J.F. co-devised the measurements, co-built the single molecule setup, prepared the samples and made all single molecule measurements. O.V. co-built the single molecule setup and co-developed analysis methodologies. G.J.H. co-devised the measurements, co-built the single molecule setup, coded the analysis methods and supervised the work. The manuscript was written by N.J.F and G.J.H.

**Competing Interests**

The authors declare no competing interests.

**Data Availability**

All relevant data is available from the authors.

**Acknowledgements**

G.J.H. acknowledges EPSRC (EP/V004921/1 and EP/V048805/1) and the Royal Society (RGS\R1\231392) for supporting this work. N.J.F.'s PhD was funded by an EPSRC DTP award (EP/T517896/1).

**References**

1. Burroughes, J. H. *et al*. Light-emitting diodes based on conjugated polymers. *Nature* **347**, 539–541 (1990).

2. Kim, Y. *et al*. A strong regioregularity effect in self-organizing conjugated polymer films and high-efficiency polythiophene:fullerene solar cells. *Nat. Mater.* **5**, 197–203 (2006).

3. He, Z. *et al*. Enhanced power-conversion efficiency in polymer solar cells using an inverted device structure. *Nat. Photonics* **6**, 591–595 (2012).

4. Fu, J. *et al*. 19.31% binary organic solar cell and low non-radiative recombination enabled by non-monotonic intermediate state transition. *Nat. Commun.* **14**, 1760 (2023).

5. Tessler, N., Denton, G. J. & Friend, R. H. Lasing from conjugated-polymer microcavities. *Nature* **382**, 695–697 (1996).

6. Heliotis, G. *et al*. Emission Characteristics and Performance Comparison of Polyfluorene Lasers with One- and Two-Dimensional Distributed Feedback. *Adv. Funct. Mater.* **14**, 91–97 (2004).

7. Samuel, I. D. W. & Turnbull, G. A. Organic Semiconductor Lasers. *Chem. Rev.* **107**, 1272–1295 (2007).

8. Clark, J. & Lanzani, G. Organic photonics for communications. *Nat. Photonics* **4**, 438–446 (2010).

9. Kuehne, A. J. C. & Gather, M. C. Organic Lasers: Recent Developments on Materials, Device Geometries, and Fabrication Techniques. *Chem. Rev.* **116**, 12823–12864 (2016).

10. Hedley, G. J., Ruseckas, A. & Samuel, I. D. W. Light Harvesting for Organic Photovoltaics. *Chem. Rev.* **117**, 796–837 (2017).

11. Mikhnenko, O. V., Blom, P. W. M. & Nguyen, T.-Q. Exciton diffusion in organic semiconductors. *Energy Environ. Sci.* **8**, 1867–1888 (2015).

12. Maniloff, E. S., Klimov, V. I. & McBranch, D. W. Intensity-dependent relaxation dynamics and the nature of the excited-state species in solid-state conducting polymers. *Phys. Rev. B* **56**, 1876–1881 (1997).

13. Dogariu, A., Vacar, D. & Heeger, A. J. Picosecond time-resolved spectroscopy of the excited state in a soluble derivative of poly(phenylene vinylene): Origin of the bimolecular decay. *Phys. Rev. B* **58**, 10218–10224 (1998).

14. Stevens, M. A., Silva, C., Russell, D. M. & Friend, R. H. Exciton dissociation mechanisms in the polymeric semiconductors poly(9,9-dioctylfluorene) and poly(9,9-dioctylfluorene-co-benzothiadiazole). *Phys. Rev. B* **63**, 165213 (2001).

15. King, S. M., Dai, D., Rothe, C. & Monkman, A. P. Exciton annihilation in a polyfluorene: Low threshold for singlet-singlet annihilation and the absence of singlet-triplet annihilation. *Phys. Rev. B* **76**, 085204 (2007).

16. Shaw, P. E., Ruseckas, A., Peet, J., Bazan, G. C. & Samuel, I. D. W. Exciton–Exciton Annihilation in Mixed-Phase Polyfluorene Films. *Adv. Funct. Mater.* **20**, 155–161 (2010).

17. Murawski, C., Leo, K. & Gather, M. C. Efficiency Roll-Off in Organic Light-Emitting Diodes. *Adv. Mater.* **25**, 6801–6827 (2013).

18. Gärtner, C., Karnutsch, C., Lemmer, U. & Pflumm, C. The influence of annihilation processes on the threshold current density of organic laser diodes. *J. Appl. Phys.* **101**, 023107 (2007).

19. Barbara, P. F., Gesquiere, A. J., Park, S.-J. & Lee, Y. J. Single-Molecule Spectroscopy of Conjugated Polymers. *Acc. Chem. Res.* **38**, 602–610 (2005).

20. Lupton, J. M. Single-Molecule Spectroscopy for Plastic Electronics: Materials Analysis from the Bottom-Up. *Adv. Mater.* **22**, 1689–1721 (2010).

21. Bolinger, J. C. *et al.* Conformation and Energy Transfer in Single Conjugated Polymers. *Acc. Chem. Res.* **45**, 1992–2001 (2012).

22. Pei, Q. & Yang. Efficient Photoluminescence and Electroluminescence from a Soluble Polyfluorene. *J. Am. Chem. Soc.* **118**, 7416–7417 (1996).

23. Grice, A. W. *et al.* High brightness and efficiency blue light-emitting polymer diodes. *Appl. Phys. Lett.* **73**, 629–631 (1998).

24. Cadby, A. J. *et al.* Film morphology and photophysics of polyfluorene. *Phys. Rev. B* **62**, 15604–15609 (2000).

25. Scherf, U. & List, E. J. W. Semiconducting Polyfluorenes—Towards Reliable Structure-Property Relationships. *Adv. Mater.* **14**, 477–487 (2002).

26. Schumacher, S. *et al.* Effect of exciton self-trapping and molecular conformation on photophysical properties of oligofluorenes. *J. Chem. Phys.* **131**, 154906 (2009).

27. Becker, K. & Lupton, J. M. Dual Species Emission from Single Polyfluorene Molecules: Signatures of Stress-Induced Planarization of Single Polymer Chains. *J. Am. Chem. Soc.* **127**, 7306–7307 (2005).

28. Da Como, E., Scheler, E., Strohriegl, P., Lupton, J. M. & Feldmann, J. Single molecule spectroscopy of oligofluorenes: how molecular length influences polymorphism. *Appl. Phys. A* **95**, 61–66 (2009).

29. Da Como, E., Borys, N. J., Strohriegl, P., Walter, M. J. & Lupton, J. M. Formation of a Defect-Free π-Electron System in Single β-Phase Polyfluorene Chains. *J. Am. Chem. Soc.* **133**, 3690–3692 (2011).

30. Adachi, T., Vogelsang, J. & Lupton, J. M. Chromophore Bending Controls Fluorescence Lifetime in Single Conjugated Polymer Chains. *J. Phys. Chem. Lett.* **5,** 2165–2170 (2014).

31. Brenlla, A. *et al.* Single-Molecule Spectroscopy of Polyfluorene Chains Reveals β-Phase Content and Phase Reversibility in Organic Solvents. *Matter* **1,** 1399–1410 (2019).

32. Wilhelm, P., Blank, D., Lupton, J. M. & Vogelsang, J. Control of Intrachain Morphology in the Formation of Polyfluorene Aggregates on the Single-Molecule Level. *ChemPhysChem* **21**, 961–965 (2020).

33. Tseng, T.-W. *et al.* Real-Time Monitoring of Formation and Dynamics of Intra- and Interchain Phases in Single Molecules of Polyfluorene. *ACS Nano* **14**, 16096–16104 (2020).

34. Brown, R. H. & Twiss, R. Q. Correlation between Photons in two Coherent Beams of Light. *Nature* **177**, 27–29 (1956).

35. Basché, Th., Moerner, W. E., Orrit, M. & Talon, H. Photon antibunching in the fluorescence of a single dye molecule trapped in a solid. *Phys. Rev. Lett.* **69**, 1516–1519 (1992).

36. Weston, K. D. *et al.* Measuring the Number of Independent Emitters in Single-Molecule Fluorescence Images and Trajectories Using Coincident Photons. *Anal. Chem.* **74**, 5342–5349 (2002).

37. Hedley, G. J. *et al.* Picosecond time-resolved photon antibunching measures nanoscale exciton motion and the true number of chromophores. *Nat. Commun.* **12**, 1327 (2021).

38. Bout, D. A. V. *et al.* Discrete Intensity Jumps and Intramolecular Electronic Energy Transfer in the Spectroscopy of Single Conjugated Polymer Molecules. *Science* **277**, 1074–1077 (1997).

39. Hu, D. *et al.* Collapse of stiff conjugated polymers with chemical defects into ordered, cylindrical conformations. *Nature* **405**, 1030–1033 (2000).

40. Huser, T., Yan, M. & Rothberg, L. J. Single chain spectroscopy of conformational dependence of conjugated polymer photophysics. *Proc. Natl. Acad. Sci.* **97**, 11187–11191 (2000).

41. Schindler, F., Lupton, J. M., Feldmann, J. & Scherf, U. A universal picture of chromophores in π-conjugated polymers derived from single-molecule spectroscopy. *Proc. Natl. Acad. Sci.* **101**, 14695–14700 (2004).

42. Pullerits, T., Mirzov, O. & Scheblykin, I. G. Conformational Fluctuations and Large Fluorescence Spectral Diffusion in Conjugated Polymer Single Chains at Low Temperatures. *J. Phys. Chem. B* **109**, 19099–19107 (2005).

43. Lin, H. *et al.* Fluorescence Blinking, Exciton Dynamics, and Energy Transfer Domains in Single Conjugated Polymer Chains. *J. Am. Chem. Soc.* **130**, 7042–7051 (2008).

44. Hedley, G. J., Steiner, F., Vogelsang, J. & Lupton, J. M. Fluctuations in the Emission Polarization and Spectrum in Single Chains of a Common Conjugated Polymer for Organic Photovoltaics. *Small* **14**, 1804312 (2018).

45. Gong, X. *et al.* Stabilized Blue Emission from Polyfluorene-Based Light-Emitting Diodes: Elimination of Fluorenone Defects. *Adv. Funct. Mater.* **13**, 325–330 (2003).

46. Romaner, L. *et al.* The Origin of Green Emission in Polyfluorene-Based Conjugated Polymers: On-Chain Defect Fluorescence. *Adv. Funct. Mater.* **13**, 597–601 (2003).

47. Kulkarni, A. P., Kong, X. & Jenekhe, S. A. Fluorenone-Containing Polyfluorenes and Oligofluorenes: Photophysics, Origin of the Green Emission and Efficient Green Electroluminescence. *J. Phys. Chem. B* **108**, 8689–8701 (2004).

48. Chang, C.-W., Sølling, T. I. & Diau, E. W.-G. Revisiting the photophysics of 9-fluorenone: Ultrafast time-resolved fluorescence and theoretical studies. *Chem. Phys. Lett.* **686**, 218–222 (2017).

49. Becker, K. *et al.* On-Chain Fluorenone Defect Emission from Single Polyfluorene Molecules in the Absence of Intermolecular Interactions. *Adv. Funct. Mater.* **16**, 364–370 (2006).

50. Honmou, Y. *et al.* Single-molecule electroluminescence and photoluminescence of polyfluorene unveils the photophysics behind the green emission band. *Nat. Commun.* **5**, 4666 (2014).

# Highly efficient exciton-exciton annihilation in single conjugated polymer chains

Nicola J. Fairbairn, Olga Vodianova and Gordon J. Hedley

School of Chemistry, University of Glasgow, Glasgow, G12 8QQ, United Kingdom

*Email: Gordon.Hedley@glasgow.ac.uk Telephone: +44 (0)141 330 1763

# Supplementary Information

**Experimental Methods**

Single molecule measurements were made on a homebuilt single molecule microscope. Excitation is provided by the 80 MHz output at 800 nm from an ultrafast optical parametric oscillator (Coherent Discovery), this is routed through a pulse picker (APE PulseSelect) to give 40 MHz, with pulses of ~120 fs duration before being frequency doubled with a BBO crystal to give 400 nm. The beam is spatially expanded with a lens pair then converted from linear to circular polarisation with a Berek Compensator (New Focus) set as a quarter waveplate for 400 nm. The beam is coupled into the microscope (Nikon Eclipse Ti2-U), reflected up to the sample with a dichroic (Thorlabs, DMLP425) before being focussed with an objective (Nikon Plan Apochromat λ, 1.45 N.A.). Incident power on the sample is controlled with a neutral density wheel before the beam is expanded, with incident powers ~3 nW at the sample.

The same objective collects emission from single polymer chains. The sample is scanned with a 2D piezo stage (Physik Instrumente, P-733.2CD), with collected emission routed back through the dichroic, through a tube lens onto a pinhole (75 µm) to remove stray light. After recollimation emission is then directed to the detection setup as described in the main text, with SPADs (MicroPhoton Devices, PD-100-CTE) in either a conventional Hanbury Brown and Twiss geometry with a 50/50 beamsplitter (Thorlabs BS013), or with an additional split between a 458 nm dichroic (Semrock, FF458-Di02) on one of the arms.

Detected photons on the SPADs are recorded with a HydraHarp 400 (PicoQuant) on separate independent channels. Overall measurement control of the piezo and photon counting is made with homecoded software (C, python and Qt framework). The data is stored as raw binary files in a first in, first out (fifo) format, containing full time tags (time with respect to last laser pulse and time with respect to start of measurement) for each photon. Fifo files are analysed in homecoded software (C, python and Qt framework), enabling extraction/analysis of intensities, decays and photon statistics.

**Sample Preparation for Single Molecule Measurements**

Samples were measured on ~170 µm thick borosilicate glass slides which were cleaned with a mixture of ultrapure water (Merck Direct-Q3) and Hellmanex III to give 2% Hellmanex solution by flushing 3 times followed by 20-minute sonication at 40°C. These steps were repeated another 2 times using ultrapure water. $N_2$ was used to dry the slides prior to exposing them to UV ozone twice for 20-minute cycles in a UV ozone cleaner (Novascan, PSD-UV8).

Single molecule PL transients were recorded from individual polymer chains of poly(9,9'-dioctylfluorene) suspended in inert thin film matrices of polymethyl methacrylate (approximately $1 \times 10^{-8}$ mg/ml). A 6% by weight 120 kDa PMMA:toluene mixture was deposited by spincoating in an $N_2$ glovebox, giving 200 nm thick films. On the microscope, single chains were measured for 30 seconds each whilst being purged with a constant flow of nitrogen to prevent degradation.

**Intensity Spot Trace**

The location of each polymer chain was determined by recording images generated by scanning the piezo over 20x20 μm regions, as shown in Figure S1.

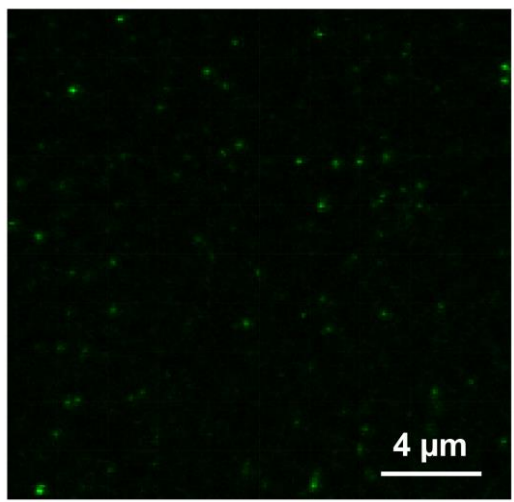

**Figure S1:** PL scan image of a 20x20 μm region of a single molecule concentration PFO sample doped in PMMA. The bright green spots are individual PFO chains.

Each polymer chain can be identified as a single diffraction-limited spot. Once all spots within an image were located, they could be measured by moving the laser to each for a full 30 seconds, producing PL transients as a function of time, as shown in Figure S2 below.

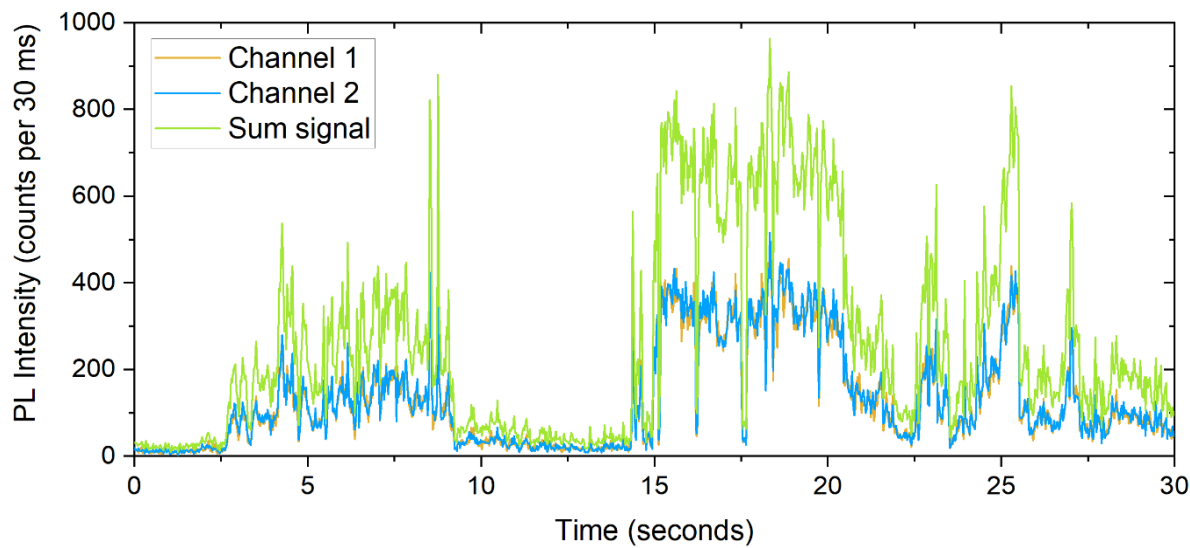

**Figure S2:** PL intensity traces of a single PFO chain measured over 30 seconds with a binning of 30 ms. Two detectors in a Hanbury Brown and Twiss geometry are used here, with each individual and the sum of them shown.

**Simulation of Expected Colour Ratio from PL Spectra**

Optical simulations were performed by taking an emission spectrum consisting of an array of wavelengths and performing spectral multiplication with all the optical components that modify this emission before it is recorded on both detectors. These components are: the microscope objective transmission, the 425 nm dichroic filter in the microscope, the non-polarising 50/50 beamsplitter, the 458 nm dichroic and the photon detection efficiency of the detector, Figure S4. This information was obtained from specification sheets of the specific components, with part numbers as listed above in the experimental methods. Both channel's detection was distinguished by reflectance or transmittance through the 458 nm dichroic mirror.

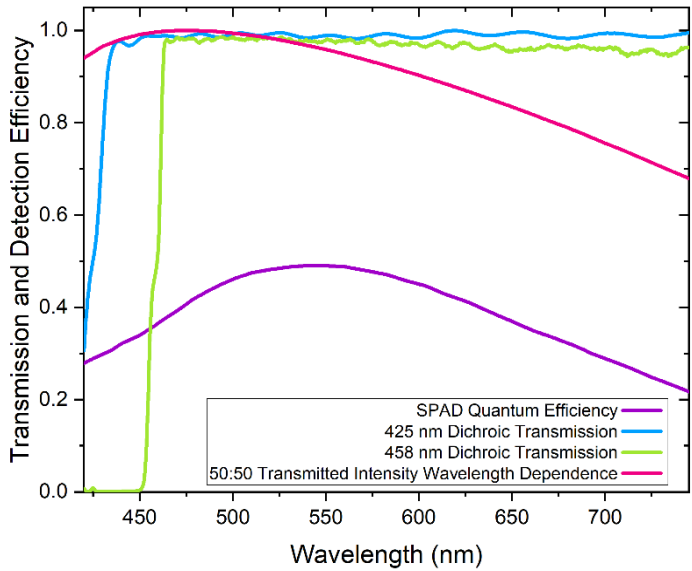

**Figure S4:** Relative efficiency of each respective optical component used for both the detection of the colour ratio and optical simulation of the emission spectra.

By multiplying these components together this gives channel intensity values detected on transmission through the 458 nm dichroic mirror onto the first detector. To calculate the channel intensity values detected on the second detector by reflectance, the 458 nm dichroic mirror component spectrum is subtracted from 1, and this 'reflectance' spectrum is multiplied together with the remaining optical component spectra.

Spectral multiplication of these curves therefore results in the final wavelength-dependence sensitivity on each detector, where the dichroic transmission corresponds to redder wavelengths (λ > 458 nm) whilst the reflectance (otherwise regarded as 1 – transmission) corresponds to bluer wavelengths (λ < 458 nm), as shown in Figure S5.

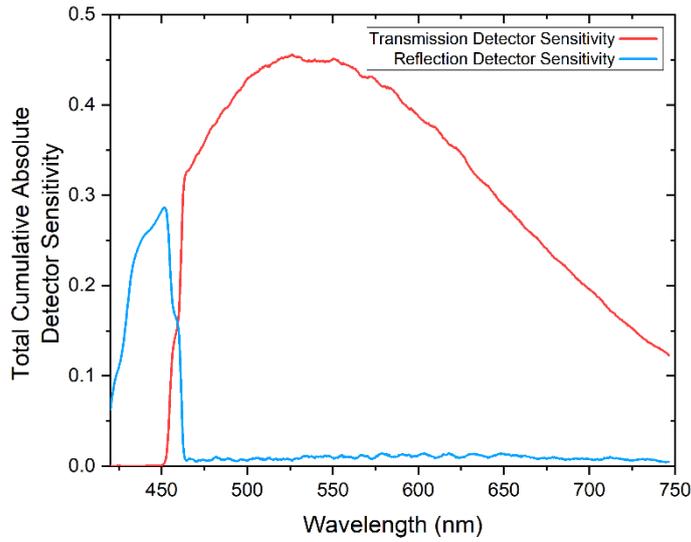

**Figure S5:** Total cumulative absolute detector sensitivity for the two detectors (one in the transmission position, one in the reflection) when incoming PL is split with the 458 nm dichroic filter.

By taking a measured PL spectrum and multiplying it by the two final SPAD sensitivity plots from Figure S5 above, we can observe the overall intensity values that would be recorded on each detector, one in transmission geometry, one in reflection. For conventional PFO emission, corresponding to steady state, this gives the two spectra shown in Figure S6 below for each detector. Integrating the two curves gives two intensity values ($I_{\lambda>458}$ and $I_{\lambda<458}$) that can be used to calculate the colour ratio, where,

$$Colour\ Ratio = \frac{I_{\lambda>458} - I_{\lambda<458}}{I_{\lambda>458} + I_{\lambda<458}}$$

For steady state PFO this gives a colour ratio of -0.24.

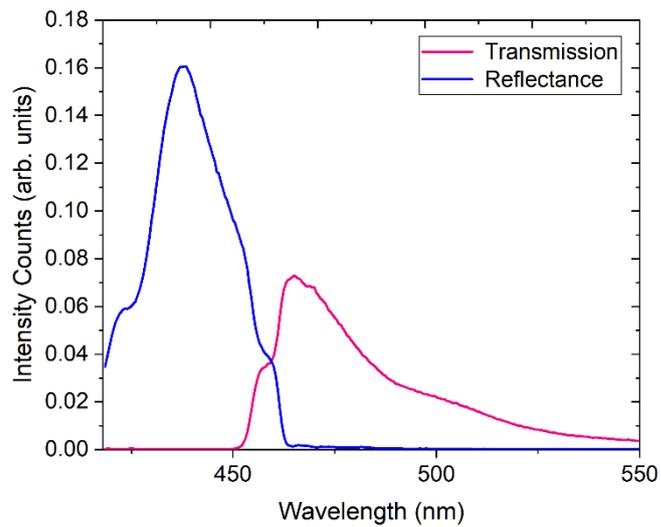

**Figure S6:** Simulated overall intensity values recorded on each SPAD for steady-state PFO emission. The two are integrated to give a calculated colour ratio of -0.24.

For beta-phase PFO, the simulated detected spectra were calculated, shown in Figure S7, giving a colour ratio of 0.008.

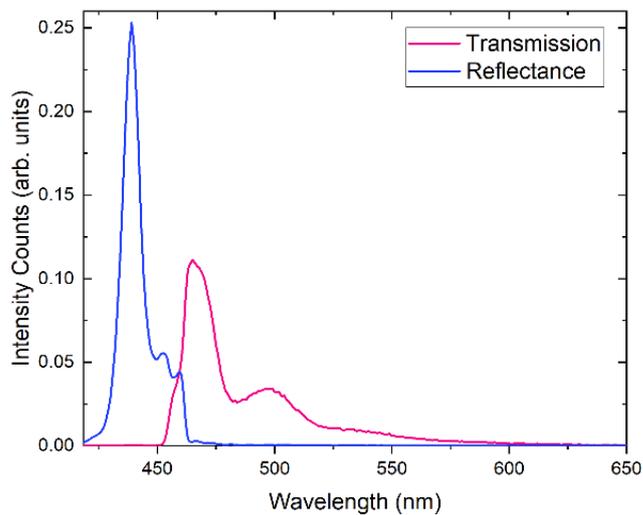

**Figure S7:** Optically simulated overall intensity values of beta-phase PFO emission recorded on each SPAD. The two are integrated to give a calculated colour ratio of 0.008.

While for fluorenone keto-defects, a PL spectrum of solely it (from reference 48 of the main text) gives the two simulated spectra shown in Figure S8 below. Here the colour ratio is calculated to be 0.945.

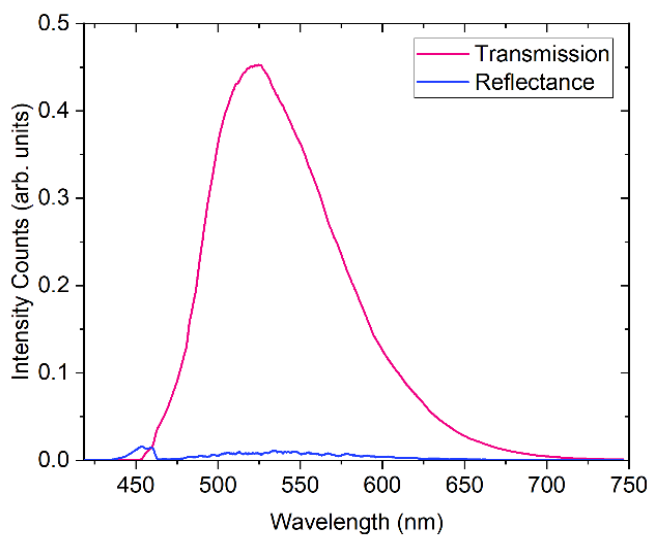

**Figure S8:** Optically simulated overall intensity values of fluorenone emission recorded on each SPAD. The two are integrated to give a calculated colour ratio of 0.945.

**Colour Ratio Binning**

Figure 3c of the main text shows measured values of the colour ratio across 5012 chains of PFO. This data is extracted with an intensity condition that at least 200 counts are present in the sum between the two colour ratio detectors (detectors 1 and 2 in Figure 3) in each 20 ms bin. All 20 ms slices that have less than 200 counts are discarded. This method is implemented in the analysis software and is done to ensure that any 20 ms periods of time when the polymer chain is emitting little or no light are not recorded in the colour ratio histogram. Additionally, this condition ensures that a high resolution of the colour ratio histogram can be shown (with a binning of 0.01), as any 20 ms slices with fewer counts would cluster around specific values, e.g. a 20 ms slice with 20 counts could only have 21 values of: -1, -0.9, -0.8 ... 0.8, 0.9, 1 and thus lead to apparent spikes in the summated distribution across all chains. This ≥ 200 count condition is also used for all subsequent antibunching/TRAB calculations, namely only 20 ms time slices with at least 200 counts are used for the results plotted in Figure 3d, as discussed below.

**Spectrally-Resolved Time-Resolved Photon Antibunching**

To calculate the time-resolved photon antibunching (Figure 3d) we have developed software routines to process the raw photon stream, extracting information of interest. The ≥ 200 count condition described above is first used to identify 20 ms time slices, and the colour ratio for them calculated. TRAB is then performed on the photons in this time slice, with the photons in channels 1 and 2 (the colour ratio arm) treated as if they come from a single new virtual channel which is correlated with those from detector 3 (the other 50% of detected photons). In effect the colour ratio arm allows us to ascertain the colour ratio for valid 20 ms slices and assign the resultant photon statistics to different spectral regions. An obvious trade-off exists when spectrally resolving photon antibunching, as the higher the spectral resolution the fewer photons will be in any correlation. Here we chose three large regions for spectral sampling, with colour ratio windows of -0.6 to 0.1, 0.2-0.7 and 0.8-1 being used.

**Comparison between 2 detector non-colour ratio and 3 detector colour ratio TRAB**

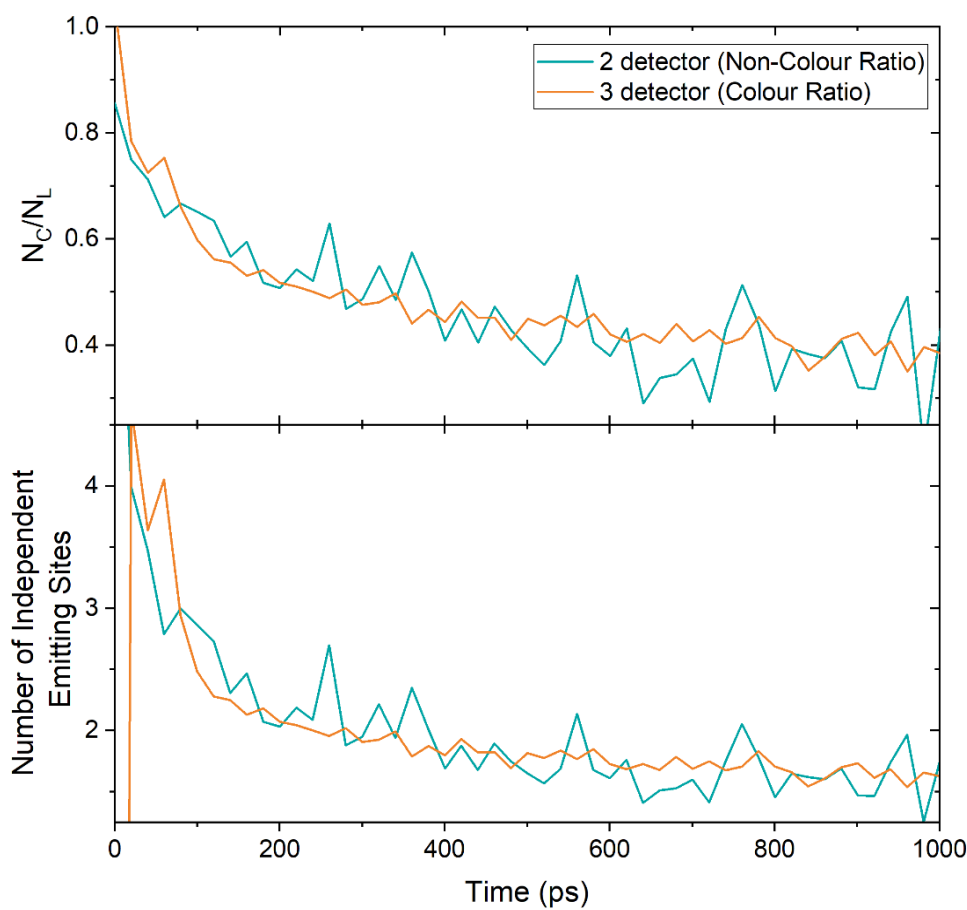

**Figure S10:** Direct comparison between the 2 detector (non-colour ratio) TRAB measurements as shown in figure 2 of approximately 2215 chains and 3 detector (colour ratio) TRAB measurements of over 5000 chains. Only the spectrally filtered TRAB are shown in figure 3 of the main text.

The direct comparison of TRAB measurements for both the non-colour ratio group of 2215 chains and the colour ratio group of 5012 chains (without any spectral filtering) is shown in Figure S10. This is plotted as both $N_C/N_L$ and as number of emitters as a function of time to enable direct comparison of the two. It is observed that they follow each other very closely, indicating that we are observing the same behaviour, regardless of the different optical setup in the three-detector measurement.